\documentclass{ws-procs9x6}

\def\be{\begin{equation}}
\def\ee{\end{equation}}
\def\bea{\begin{eqnarray}}
\def\eea{\end{eqnarray}}
\def\ft#1#2{{\textstyle{{\scriptstyle #1}\over {\scriptstyle #2}}}}
\def\dalemb#1#2{{\vbox{\hrule height .#2pt
        \hbox{\vrule width.#2pt height#1pt \kern#1pt
                \vrule width.#2pt}
        \hrule height.#2pt}}}
\def\square{\mathord{\dalemb{6}{7}\hbox{\hskip1pt}}}
\def\oneone{\rlap 1\mkern4mu{\rm l}}
\def\R{{\mathbb R}}

\begin{document}

\title{Counterterms, Holonomy and Supersymmetry}

\author{K.S. Stelle\footnote{\uppercase{R}esearch
supported in part by the \uppercase{EC} under 
\uppercase{RTN} contract \uppercase{HPRN-CT}-2000-00131 and by \uppercase{PPARC}
under rolling grant \uppercase{PPA/G/O}/2002/00474.}}

\address{Theoretical Physics Group,\\
Imperial College London,\\
Prince Consort Road,\\
London SW7 2AZ, UK\\ 
E-mail: k.stelle@imperial.ac.uk}

\maketitle

\abstracts{
The divergence structure of supergravity has long been a topic of concern because
of the theory's non-renormalizability. In the context of string
theory, where perturbative finiteness should be achieved, the supergravity
counterterm structures remain nonetheless of importance because they still occur,
albeit with finite coefficients. The leading nonvanishing supergravity counterterms have a
particularly rich structure that has a bearing on the preservation of
supersymmetry in string vacua in the presence of perturbative string
corrections. Although the holonomy of such manifolds is
deformed by the corrections, a Killing spinor structure nevertheless can persist.
The integrability conditions for the existence of such Killing spinors remarkably remain
consistent with the perturbed effective field equations.}

\section{Supergravity Counterterms}

The ultraviolet divergences of quantized general relativity and its various
matter couplings have posed a key problem for the reconciliation of quantum
mechanics and relativity. The potential for ultraviolet trouble with gravity was apparent
already since the 1930's from rudimentary power counting, in
consequence of Newton's constant having dimensionality $[{\rm length}]^2$. When
detailed calculations of gravitational Feynman diagrams became possible in the
1970's\cite{tH-V}, this became a reality with the first
calculations of divergence structures that are not present in the original second-order
action. As ever, in the key issues involving gravity and its quantization, Stanley
Deser played a major r\^ole in this development\cite{brandeis1,brandeis2}.

As disastrous as the ultraviolet problem was for quantized field theories containing gravity,
there was nonetheless some hope that a clever combination of fields might save the day by arranging for the divergences to
marvelously cancel. The prime candidate for an organizing principle that might engineer this was supersymmetry, and
when supergravity came forth\cite{supergravity}, there was palpable hope that it might enable the
construction of some jewel-like theory that could resolve (maybe uniquely) the ultraviolet problems.
This hope was encouraged by the development of  non-renormalization theorems for chiral supermatter\cite{nonren} and by initial
calculations showing that supergravity also had better-than-generic ultraviolet behavior. For one-loop Feynman diagrams, the
divergences cancel in pure $N=1$ supersymmetry, as one can see by summing the contributions of the different field species occurring
in the loop. A range of differing arguments was advanced on formal grounds to why these
cancellations occur and why they could be expected to persist at the two-loop level (despite the
prohibitive difficulty of actually performing such calculations). One approach\cite{helicity} that has much
current resonance focused on helicity conservation properties. 

In many of these early developments, the lively scientific atmosphere at Brandeis yielded important understanding of
these ultraviolet problems. For me, as a graduate student there at the time, it was a marvellous training ground for
learning the way physics should really be done, but one with a decidedly European flavor. Stan Deser was without
doubt the leader in these matters, and it summons pleasant recollections to think back to how these fundamental
issues were grappled with. Given Stan's status as {\it doyen} of the canonical formalism, another natural
development we got into at the time was the canonical formulation of supergravity.\cite{dkscan} Although not
directly related to the issue of infinities, this revealed a number of essential duality properties of the theory
and it also provides, {\it via} the duality-related form of the constraints, a link to the Ashtekar variable program
for quantum gravity.

The clearest reason for ultraviolet cancellations was the requirement that the counterterms preserve local
supersymmetry. This was given a clear expression in the detailed analysis of $N=1$ supergravity counterterms that we
performed together with Stanley and John Kay in Ref.\cite{dks} The result was not ultimately encouraging for the
prospects of finiteness, but it was intriguing nonetheless. The first relevant $N=1$ supergravity counterterm occurs
at the three-loop level, at which order power counting leads one to expect an expression quartic in curvatures, since
at one loop the leading logarithmic divergences are of fourth order in derivatives and each loop adds two more to this
count. However, at one and two loops, the possible counterterm structures happen to vanish subject to the classical
field equations, so they can be eliminated by field redefinitions renormalizations.

The most intriguing aspect of the three-loop counterterm was its geometrical structure: the
purely gravitational part is the contracted square of the Bel-Robinson tensor\cite{BR}
$T_{\mu\nu\alpha\beta}=-R^\lambda{}_\alpha{}^\rho{}_\mu
R_{\lambda\beta\rho\nu}+{}^*R^\lambda{}_\alpha{}^\rho_\mu {}^*R_{\lambda\beta\rho\nu}$. Subject to
the Einstein field equations, this tensor is covariantly divergence-free on any index, totally
symmetric and totally traceless. Thus, it is a higher-order analogue of the stress tensor, whose
contracted square occurs in the the (nonrenormalizable) one-loop divergences of the gravity plus
Yang-Mills system\cite{brandeis2}. Similarly, in $N=1$ supergravity plus super Yang-Mills, one
encounters the stress-tensor supermultiplet $(T_{\mu\nu}, J_{\mu\,\alpha}, C_\mu)$, where
$J_{\mu\,\alpha}$ is the matter supersymmetry current and $C_\mu$ is the matter axial current. These
come together in the counterterm $\int d^4x (T^{\mu\nu}T_{\mu\nu} + {\rm i}\bar
J^\mu\gamma^\rho\partial_\rho J_\mu - \ft32 C^\mu\square C_\mu)$.

In extended supergravities, the gravitational and lower-spin contributions give expressions that manage to vanish
subject to the classical field equations at the one- and two-loop levels, as Stanley and John Kay found already in
the $N=2$ case.\cite{dk} In the early days, it was hoped that this situation might continue on
to higher orders, but the added constraints of local supersymmetry (and this for all degrees of extension) prove to
be exhausted at the next, three-loop, order.  The corresponding $N=1$ counterterm, whose structure continues to
figure importantly in quantum gravity discussions in the string era, is a natural generalization of the one-loop
matter divergence structure:
\bea
\Delta_3 I &=& \int d^4x
\Big((T^{\mu\nu\alpha\beta}+H^{\mu\nu\alpha\beta})(T_{\mu\nu\alpha\beta}+H_{\mu\nu\alpha\beta})\cr
&& \qquad\qquad\qquad + {\rm i} \bar
J^{\mu\alpha\beta}\gamma^\rho\partial_\rho J_{\mu\alpha\beta} -
\ft32 C^{\mu\alpha\beta}\,\square C_{\mu\alpha\beta}\Big)\label{BR2}
\eea
where $H_{\mu\nu\alpha\beta}=-\ft{\rm i}2\bar
f^\lambda{}_\alpha(\gamma_\mu\partial_\nu+\gamma_\nu\partial_\mu)f_{\beta\lambda}$,
$J_{\mu\alpha\beta}=\ft12 R^\lambda{}_{\alpha\rho\tau}\sigma^{\rho\tau}\gamma_\mu f_{\lambda\beta}$,
$C_{\mu\alpha\beta}=-\ft{\rm i}2\bar f^\lambda{}_\alpha\gamma_5\gamma_\mu f_{\beta\lambda}$ in which
$f_{\alpha\beta}=\partial_\alpha\psi_\beta-\partial_\beta\psi_\alpha$ is the Rarita-Schwinger field strength for the
gravitino field. $H_{\mu\nu\alpha\beta}$ plays the role of the 'matter' contribution to the Bel-Robinson `stress'.
while $J_{\mu\alpha\beta}$ and $C_{\mu\alpha\beta}$ are similarly higher-order analogues of the supersymmetry current
$J_\mu$ and axial current $C_\mu$.

Although direct Feynman diagram calculations of the divergent coefficients of such higher-loop counterterms remain out of reach,
other techniques for evaluating such divergences have progressed immensely since the late 1970's. Clever use of unitarity cutting rules
plus dimensional regularization\cite{bddpr} have yielded the result that analogues of the $D=4$ three-loop counterterm do
indeed occur with an infinite coefficient for all $N\le6$ extended supergravities, but that the
$N=8$ theory (which is the same as $N=7$) manages to remain finite until it too succumbs at five loops (by which time the
ordinary Feynman diagram approach would involve something like $10^{30}$ terms \ldots). Similar considerations also apply to supergravity
divergences in higher dimensions, where the corresponding divergences occur at lower loop orders, {\it e.g.}\ in $D=11$ one has
divergences already at the two loop level. In the nonlocal parts of the four-graviton amplitude in such cases one again finds
analogues of the $({\rm Bel-Robinson})^2$ counterterm.\cite{dsem} 

The special circumstance of the $N=8$ theory remains highly intriguing. It is likely to be
analogous to that which obtains for maximal ({\it i.e.}\ 16-supercharge, corresponding to $N=4$ in $D=4$) super Yang-Mills theory in 5 or 6
dimensions, which also becomes nonrenormalizable, but later than na\"{\i}vely expected. The SYM divergence onset can be understood\cite{hs} using
superspace power counting\cite{griegelcek,miraculous} together with the existence of a 12-supercharge harmonic superspace formalism\cite{gikos}.
Although the story remains incomplete, it may be anticipated that something similar is going on in maximal supergravity, perhaps with an $N=6$
harmonic superspace formalism.

\section{String corrections}

String theory may be viewed as a `physical' regulator for the divergent supergravity theories. Instead of a
Feynman integral  cutoff, one has the string length $\sqrt{\alpha'}$. Counterterms that would have occurred
with divergent coefficients in a supergravity field theory now occur with finite $\alpha'$-dependent
coefficients in quantum string corrections. In particular, the $({\rm curvature})^4$ counterterms of
$D=4$, 3-loop supergravity now are present at the $\alpha'\,^3$ level in superstring theories. Generally, these
string corrections have been calculated using the Neveu-Schwarz-Ramond (NSR) formalism in string light-cone
gauge. As in the analogous field theory case, the first corrections not vanishing subject to the
classical supergravity field equations (which are now removable by nonlinear redefinitions of the background fields
that the string propagates on) occur at the $\alpha'\,^3$ level. One may write the string tree-level correction in the
general form
\be
\Delta I = \xi\alpha'\,^3\int d^{10}x\sqrt{-g}\,e^{-2\phi}\,Y \label{countform}
\ee
where the dependence on the dilaton $\phi$ is appropriate to string tree level. A similar form is obtained
at one string loop, but without any $e^\phi$ factor, as is appropriate for one-loop order in string
perturbation theory. 

From the string light-cone gauge calculations, the integrand $Y$ in (\ref{countform}) can be written in
terms of a Berezin integral over an anticommuting spinor field $\psi_{L,R}$. The $({\rm curvature})^4$
correction thus takes the form\cite{grosswitten,freemanpope}
\be
Y = \int d^8\psi_Ld^8\psi_R\exp(\bar\psi_L\Gamma^{ij}\psi_LR_{ijkl}\bar\psi_R\Gamma^{kl}\psi_R)\label{Yform}
\ee
where $i,j=1,\ldots 8$ are light-cone transverse indices,
$\Gamma^{ij}=\ft12(\Gamma^i\Gamma^j-\Gamma^j\Gamma^i)$ are
${\rm SO}(8)$ Gamma matrices and $\psi_L$ and $\psi_R$ are left- and right-handed ${\rm SO}(8)$ chiral
spinors. From the fact that Berezin integration gives zero except when a linear expression in each spinor field
is integrated over, one sees immediately that (\ref{Yform}) produces exclusively $({\rm curvature})^4$
corrections. 

Letting $\alpha$ and $\dot\alpha$ be 8-valued $R$, $L$ spinor indices, one has, up to a proportionality
constant, 
\bea
Y &=& \epsilon^{\alpha_1\alpha_2\cdots\alpha_8}\epsilon^{\dot\beta_1\dot\beta_2\ldots\dot\beta_8}
\Gamma_{\alpha_1\alpha_2}^{i_1i_2}\cdots\Gamma_{\alpha_7\alpha_8}^{i_7i_8}
\Gamma_{\dot\beta_1\dot\beta_2}^{j_1j_2}\cdots\Gamma_{\dot\beta_7\dot\beta_8}^{j_7j_8}\cr
&&\qquad\qquad \times R_{i_1i_2j_1j_2}R_{i_3i_4j_3j_4}R_{i_5i_6j_5j_6}R_{i_7i_8j_7j_8}\ .
\eea
Working this out in more detail, one finds $Y=Y_0-Y_2$, where
\bea
Y_0=\ft1{64}t^{i_1\ldots i_8}t^{j_1\ldots
j_8}R_{i_1i_2j_1j_2}R_{i_3i_4j_3j_4}R_{i_5i_6j_5j_6}R_{i_7i_8j_7j_8}\cr 
Y_2=\ft1{256}\epsilon^{i_1\ldots i_8}\epsilon^{j_1\ldots
j_8}R_{i_1i_2j_1j_2}R_{i_3i_4j_3j_4}R_{i_5i_6j_5j_6}R_{i_7i_8j_7j_8}
\eea
in which $t^{i_1\ldots i_8}$ is defined by
\be
t^{i_1\ldots i_8}M_{i_1i_2}\ldots M_{i_7i_8}=24M_i{}^jM_j{}^kM_k{}^\ell M_\ell{}^i-6(M_i{}^jM_j{}^i)^2
 .
\ee

In making a light-cone gauge choice for the string variables in order to derive the form of these 
$({\rm curvature})^4$ corrections, one has in fact to restrict the background curvature to the transverse 8
coordinates $i_1\ldots i_8$, so in fact the term $Y_2$ contributes a total derivative here, since it becomes the
Euler density in $D=8$. Since the string tree-level correction (\ref{countform}) multiplies this by
$e^{-2\phi}$, there still is a contribution to the Einstein equation, but this becomes proportional to
$\partial\phi$, so it vanishes if one is considering corrections to background field solutions that have an
initially constant dilaton $\phi$. The $D=8$ Euler density $Y_2$ comes in very usefully in the equation for
the dilaton itself, since the combination $Y_0-Y_2$ actually vanishes for all spaces that are endowed with a Killing
spinor. In consequence, at order $\alpha'\,^3$, the dilaton's contribution to the correction for supersymmetric spaces
can be integrated out explicitly, leaving one with an expression that is purely gravitational, and
which is a direct generalization of the $D=4$ $(\hbox{Bel-Robinson})^2$ supergravity counterterm (\ref{BR2}).

For spaces with suitable initial supersymmetry (so that there is at least one holonomy singlet among the spinors
coupling to `front' and rear' indices of the curvature in the exponent of (\ref{Yform})), the variation of the quantum
correction (\ref{countform}) simplifies further in that the `explicit' metric variations also vanish.\cite{lpst2} This
is the case for $D\le 8$ spaces with special holonomies such as ${\rm SU}(3)$, ${\rm SU}(4)$, $G_2$ or ${\rm Spin}_7$.
Consequently, the contributions to the Einstein equations arise purely from the `implicit' metric variations coming
from the spin connections. One obtains in each case a corrected Einstein equation that to order
$\alpha'\,^3$ becomes\cite{lupost,g2mod,lpst2}
\be
R_{ij}+2\nabla_i\nabla_j\phi-\alpha'\,^3X_{ij}=0\label{graveqn}
\ee
in which the correction $X_{ij}$ arises from the connection variations, giving a correction of the form
\be
X_{ij}=\nabla^k\nabla^\ell X_{ikj\ell}
\ee
where $X_{ikj\ell}$ is an expression cubic in curvatures with symmetries similar to those of the curvature tensor:
$[ik]$, $[jl]$ antisymmetric but $[ik]\leftrightarrow[j\ell]$ symmetric under pair interchange.
Tracing the corrected gravitational equation and combining it with the dilaton equation one obtains to this order
\be
2\,\square\phi+\alpha'\,^3X=0\ ,
\ee
where the $X=g^{ij}X_{ij}$ correction arises purely from the gravitational equation trace, since the dilaton equation
itself does not have order $\alpha'\,^3$ corrections for initially supersymmetric spaces, as we have seen above.
Moreover, for the special holonomy manifolds in question, one finds
\be
g^{ij}X_{ikj\ell}=g_{k\ell}Z\ .
\ee
Thus, $X=\square Z$ and consequently one can solve explicitly for the dilaton correction: for $\phi={\rm
const}+\phi_1$, where
$\phi_1$ is the correction to the initially constant dilaton. One finds $\phi_1=-\ft12\alpha'\,^3 Z$ so the corrected
Einstein equation becomes
\be
R_{ij}=\alpha'\,^3(\nabla_i\nabla_j Z+\nabla^k\nabla^\ell X_{ikj\ell})\ .\label{correqform}
\ee

\section{Special Holonomy}

To see how the corrected form (\ref{correqform}) of the Einstein equation influences the background
field solutions that initially have special holonomy, consider first the case of spaces with structure $M_8=\R\times
K_7$ where $K_7$ is, at order $\alpha'\,^0$, a 7-manifold with holonomy $G_2$. Similar conclusions are obtained for
8-manifolds of ${\rm Spin}_7$ holonomy.\cite{lpst2} To study the $G_2$ case, pick the following basis for the
${\rm SO}(8)$ Dirac $\Gamma$ matrices:
\be\label{gammarep}
\tilde\Gamma^{\underline i} 
= \sigma_2 \otimes \Gamma^{\underline {i}}  \qquad {\underline 
i}=1,\dots,7; \qquad 
\tilde\Gamma^8 = -\sigma_1 \otimes \oneone_8\ ,
\ee
where the $\Gamma^{\underline {i}}$ are antisymmetric imaginary $8\times 8$ ${\rm SO}(7)$ $\Gamma$-matrices; signs are
chosen such that $i\Gamma^{\underline 1}\cdots \Gamma^{\underline 7} = \oneone_8$. Chiral ${\rm SO}(8)$ spinors are
eigenspinors of $\tilde\Gamma_9 \equiv \tilde\Gamma^{\underline 1}\cdots\tilde\Gamma^{\underline 8} =  
\sigma_3\otimes \oneone_8$, so $\Psi=\begin{pmatrix}\Psi_+\\ \Psi_-\end{pmatrix}$ where $\Psi_+$ and $\Psi_-$ are real
8-component ${\rm SO}(7)$ spinors. Consequently, for manifolds of $G_2\subset {\rm SO}(7)$ holonomy, the $8_\pm$
representations decompose as $8_\pm\rightarrow 7\oplus 1$. Accordingly, the $({\rm curvature})^4$ correction
\be
Y \propto \int d^8\psi _+ d^8\psi_- \, 
\exp\left[ \left(\bar\psi_+\, \Gamma^{ij}_+\,
\psi_+\right)\left(\bar\psi_-\,  \Gamma^{k\ell}_-\, 
\psi_-\right) R_{ijk\ell}\right]
\label{fermionform2}
\ee
satisfies the requirements for vanishing of `explicit' metric variations in (\ref{Yform}), and the resulting
corrections to the Einstein equations arising solely from the connection variations are of the form
(\ref{correqform}).

The value of $Y$ in (\ref{fermionform2}) is zero for manifolds of initial $G_2$ holonomy ({\it i.e.}\ before the effects of
$\alpha'$ corrections are included), owing to the presence of the holonomy singlets in both the $\Psi_\pm$ decompositions, together with the
rules of Berezin integration, which give a vanishing result for $\int d\theta$ integrals without a corresponding
$\theta$ in the integrand. This accounts for the absence of direct $\alpha'\,^3$ corrections to the dilaton equation,
as we have noted. 

The vanishing of $Y$ for such spaces does not, however, imply the vanishing of its full variation. This has to be
performed without restriction to spaces of any particular holonomy, although the initial holonomy is subsequently used
in evaluating the result after variation. The only surviving terms in the variation of (\ref{fermionform2}) are those
where the singlets in the $8_\pm$ decompositions go onto the `front' and `back' of the same {\em varied} curvature,
since the only way one can get a nonvanishing result is to keep the singlet products from contracting with unvaried
curvatures. This observation gives a way to write the variation in a nice fashion (where now $i,j = 1,\ldots,7$):
\be
\delta Y \propto  \epsilon^{m i_1\cdots i_6}\, \epsilon^{n j_1\cdots j_6} \, 
R_{i_1 i_2 j_1 j_2}\ R_{i_3i_4j_3j_4}\  R_{i_5 i_6 j_5 j_6}\, 
     c^{ij}{}_{m}\, c^{k\ell}{}_{n}\ \nabla_i\nabla_k \, \delta 
g_{j\ell}\ ,\label{deltaY}
\ee
where $c_{ijk} = {\rm i}\, \bar\eta\, \Gamma_{ijk}\, \eta$ is the covariantly constant 3-form that characterizes a $G_2$ holonomy manifold. 

The variation (\ref{deltaY}) thus takes the general form (\ref{correqform}) with 
\be
X_{ijk\ell} = c_{ik m}\, c_{j\ell n}\, Z^{mn}\ ,
\label{Xdef}
\ee
where 
\be
Z^{mn} \equiv  \ft1{32} \epsilon^{m i_1\cdots i_6}\, \epsilon^{n j_1\cdots j_6} 
\, 
R_{i_1 i_2 j_1 j_2}\cdots R_{i_5 i_6 j_5 j_6}\ ,\quad Z=g_{mn}Z^{mn}\ .\label{zmndef}
\ee
{\it i.e.}\ the corrected Einstein equation is now
\be\label{correctedg2} 
R_{ij} = c\alpha'\,^3 \left[\nabla_i\nabla_j Z + c_{ikm}c_{j\ell n}
\nabla^k\nabla^\ell Z^{mn}\right]\ .
\ee 

The corrected Einstein equation (\ref{correctedg2}) modifies the curvature at order $\alpha'\,^3$ so as to give an
apparently {\em generic} ${\rm SO}(7)$ holonomy, {\it i.e.}\ the initial $G_2$ special holonomy is lost as a result
of the
$\alpha'$ corrections. To see this, note that the integrability condition for the existence of an ordinary Killing
spinor $\eta$ satisfying $\nabla_i\eta=0$ is $R_{ijk\ell}\, c^{k\ell}{}_{mn} = 2 R_{ijmn}$ where
$c^{ijk\ell} \equiv \ft16 \epsilon^{ijk\ell mnp}\, c_{mnp} = \bar\eta\,
\Gamma_{ijk\ell}\, \eta$ is the Hodge dual of $c_{ijk}$ in $D=7$. Taking the trace of this integrability condition,
one finds that $G_2$ holonomy {\rm requires} Ricci flatness, so the corrected Ricci tensor (\ref{correctedg2})
definitely takes the metric out of the class of $G_2$ holonomy manifolds.

This should be contrasted with the more familiar case of K\"ahler manifolds, where the
corrected Ricci form is required to be a cohomologically trivial (1,1) form, but is not
required to vanish. One may see this explicitly is by considering an initial 7-manifold
$K_7=\R\times K_6$, where at order $\alpha'\,^0$, $K_6$ is K\"ahler and Ricci flat. This
fits into the above $G_2$ holonomy discussion when one recognises that the only non-zero
component of $Z_{mn}$ in this case is $Z_{77}=Z$, while $c_{ij7}=J_{ij}$ is the K\"ahler
2-form. In this K\"ahler case, the corrected Einstein equation becomes
\be
R_{ij} = 
c\alpha'\,^3 \left[\nabla_i\nabla_j  + J_{ik}J_{j\ell} 
\nabla^k\nabla^\ell\right] Z\ .
\label{kahlercase}
\ee
Going over to a Darboux complex coordinate basis $i,j=1,\ldots,6\rightarrow a,\bar
a=1,2,3$, one then has the standard Calabi-Yau result
\be
R_{a\bar b}=c\alpha'\,^3\nabla_a\nabla_{\bar b} Z\ ,\label{Ricciform}
\ee
which is a cohomologically trivial (1,1) form, but which does not destroy the K\"ahler
structure (which depends on the vanishing of the Nijenhuis tensor, which is not
disturbed).

\section{Corrected Killing equations}

Despite the fact that the string $\alpha'$ corrections perturb manifolds of initially
special holonomy into manifolds of generic Riemannian holonomy, another remarkable property
of the Bel-Robinson-descendant string corrections is that a manifold's initial supersymmetry
can nonetheless be preserved. This can happen because the Killing spinor equation 
can itself be modified in such a way that its corrected integrability condition reproduces precisely
the corrected Einstein equations. To see how this can happen, seek a condition
$\hat\nabla_i\eta=0$, where $\hat\nabla_i=\nabla_i+\alpha'\,^3Q_i$. We need to choose
$Q_i$ such that the integrability condition $[\hat\nabla_i,\hat\nabla_j]\, \eta=0$ yields
the corrected Einstein equation. 

One has directly the integrability condition
\be\label{intcon}
\ft14 R_{ijk\ell}\, \Gamma^{k\ell}\, \eta + c\alpha'\,^3 Q_{ij}\, \eta =0\ ,
\ee
where
\be
Q_{ij}\equiv \nabla_i\, Q_j - \nabla_j\, Q_i\ . 
\ee

In the case of a manifold of initial $G_2$ holonomy, one can use the Fierz identity
$\Gamma_i\, \eta\, \bar\eta\, \Gamma_i + \eta\, \bar\eta = \oneone$ to find
\be
R_{ijk\ell}\, c^{k\ell}{}_{m} + 4c\alpha'\,^3  {\rm i}\, \bar\eta\, \Gamma_m\, 
Q_{ij}\, \eta =0\ ,\label{intcon2}
\ee
where $\bar\eta Q_{ij}\ \eta =0$. Multiplying by $\Gamma_i$ and using the Fierz identity,
one obtains a supersymmetry integrability condition involving the corrected Ricci tensor
\be
R_{ij} = 2c\alpha'\,^3\,  \bar\eta\, \Gamma_{(jk}\, Q_{i)}{}^k\, 
\eta\ .\label{intcon3}
\ee
The condition (\ref{intcon3}) must then be consistent with the corrected Einstein equations
for some choice of the Killing spinor correction $Q_i$.

In principal, one should be able to find the Killing spinor correction $Q_i$ by an
exhaustive study of the supersymmetry properties of the $({\rm curvature})^4$ counterterm.
This would require first determining the structure of the superpartners to the pure $({\rm
curvature})^4$ part by varying it subject to the original $\alpha'\,^0$ supersymmetry
transformations but subject to the $\alpha'\,^0$ field equations, then relaxing the latter and calculating
the required corrections to the gravitino supersymmetry transformation. This is a long process which has
not been carried out for the maximal $D=11$ and $D=10$ supergravities. However, the requirements
for $Q_i$ nonetheless allow one to find out its structure. The answer, {\it i.e.}\ the solution to (\ref{intcon}) is
\be
Q_i = -\ft{\rm i}{2}\,  c_{ijk}\, \nabla^j\, Z^{k\ell}\, \Gamma_\ell\ .
\label{Qidef}
\ee
The integrability condition for the modified Killing spinor condition then reproduces precisely the
corrected Einstein equation (\ref{correctedg2}).

The Killing spinor correction (\ref{Qidef}) seemingly depends on special properties of the order
$\alpha'\,^0$ manifold, since it involves the $G_2$ manifold's covariantly constant 3-form $C_{ijk}$.
However, another remarkable structural feature emerges here. The Killing spinor correction
(\ref{Qidef}) can be rewritten in a form that does not make use of any special tensors on the manifold:
\be
Q_i = -\ft34 \,  (\nabla^j\, R_{ik m_1 m_2})\, R_{j\ell m_3 m_4}\, 
R^{k\ell}{}_{m_5 m_6}\, \Gamma^{m_1 \cdots m_6}\ .\label{universalQ}
\ee
Moreover, this is precisely the same expression as one finds from the study of corrections to $D=6$
K\"ahler manifolds\cite{cfpss}, so there is a strong argument for the {\em universality} of the result
(\ref{universalQ}).

\section{Conclusion}

The quantum field theoretic approach to quantum gravity, of which Stan Deser is a key pioneer continues to
yield important insights into a theory of which we still have only glimpses. The main approach to quantum gravity 
has changed from the canonical formulation to supergravity and on to superstrings, but there is considerable continuity in
certain central elements of the story. The ultraviolet problem for gravity, which gave rise to much soul-searching
about the nature of the entire perturbative quantum gravity program, has now more or less been solved. Accordingly, one can
now begin to actually look at the perturbatively finite theory that lies behind. Despite the evolution in dynamical details, the
analytical approach of focusing on symmetries and their consequences remains an important strategy.

In the examples that we have looked at, ultraviolet counterterms that spelled the end of supergravity as a fundamental
theory in its own right remain nonetheless of keen interest as finite local contributions to the supergravity
effective action for superstrings or M-theory. They have a set of `miraculous' properties that appear to make them
precisely tailored to preserving the integrity of the underlying, still incompletely known, string or M-theory. In
particular, they may lead to important insights into the structure of M-theory, for which we still have no full
microscopic formulation. 

An example of this is the link between the $C_{[3]}\wedge R^4$ coupling in M-theory and the
$R^4$ terms that arise in type IIA string theory at the one loop level, which are in turn related to M-theory $({\rm
curvature})^4$ terms by dimensional oxidation. These two types of terms are related\cite{greenetal} by on-shell
supersymmetry; the relation is also crucial\cite{lpst2} for the way in which the supersymmetry of an initially ${\rm
SU}(5)$ holonomy K\"ahler manifold can be preserved despite the fact that the $({\rm curvature})^4$ corrections in this
case destroy the K\"ahler structure, yielding a general complex $D=10$ manifold. {\it Via} a sequence of `miracles'
analogous to those we have sketched here for the string tree level $G_2$ case, the initial supersymmetry of such a
background turns out to be preserved thanks to interrelated corrections to the Einstein and 4-form field equations. The
$C_{[3]}\wedge R^4$ terms play a key r\^ole\cite{becbec} in this mechanism, because they force the turning on of a necessary
amount of 4-form flux. The same terms are also crucial for the elimination of the sigma-model anomalies of the M5
brane\cite{duffliuminasian} and for duality between M2 and M5 branes. The quartic curvature corrections thus are deeply
related to the internal consistency of our best chance for a fundamental theory of quantum gravity.

\end{document}